**Visualization of the flat electronic band in twisted bilayer graphene near the magic angle twist**

M. Iqbal Bakti Utama[1,2,3,†], Roland J. Koch[4,†], Kyunghoon Lee[1,3,†], Nicolas Leconte[5], Hongyuan Li[1,6], Sihan Zhao[1], Lili Jiang[1], Jiayi Zhu[1], Kenji Watanabe[7], Takashi Taniguchi[7], Paul D. Ashby[8], Alexander Weber-Bargioni[8], Alex Zettl[1,3,9], Chris Jozwiak[4], Jeil Jung[5,*], Eli Rotenberg[4], Aaron Bostwick[4,*], Feng Wang[1,3,9,*]

1. Department of Physics, University of California at Berkeley, Berkeley, California 94720, USA
2. Department of Materials Science and Engineering, University of California at Berkeley, Berkeley, California 94720, USA
3. Materials Sciences Division, Lawrence Berkeley National Laboratory, Berkeley, California, 94720, USA
4. Advanced Light Source, Lawrence Berkeley National Laboratory, Berkeley, California, 94720, USA
5. Department of Physics, University of Seoul, Seoul, South Korea
6. Graduate Group in Applied Science and Technology, University of California at Berkeley, Berkeley, California 94720, USA
7. National Institute for Materials Science, 1-1 Namiki, Tsukuba, 305-0044, Japan
8. Molecular Foundry, Lawrence Berkeley National Laboratory, Berkeley, California, 94720, USA
9. Kavli Energy NanoSciences Institute at the University of California, Berkeley and the Lawrence Berkeley National Laboratory, Berkeley, California 94720, United States

† These authors contributed equally to this work
* e-mail: fengwang76@berkeley.edu, abostwick@lbl.gov, jeiljung@uos.ac.kr

**Bilayer graphene was theorized to host a moiré miniband with flat dispersion if the layers are stacked at specific twist angles known as the "magic angles"[1,2]. Recently, such twisted bilayer graphene (tBLG) with the first magic angle twist was reported to exhibit correlated insulating state and superconductivity[3,4], where the presence of the flat miniband in the system is thought to be essential for the emergence of these ordered phases in the transport measurements. Tunneling spectroscopy[5-9] and electronic compressibility measurements[10] in tBLG have revealed a van Hove singularity that is consistent with the presence of the flat miniband. However, a direct observation of the flat dispersion in the momentum-space of such moiré miniband in tBLG is still elusive. Here, we report the visualization of the flat moiré miniband by using angle-resolved photoemission spectroscopy with nanoscale resolution (nanoARPES). The high spatial resolution in nanoARPES enabled the measurement of the local electronic structure of the tBLG. We clearly demonstrate the existence of the flat moiré band near the charge neutrality for tBLG close to the magic angle at room temperature.**

An implication of an electronic band with flat momentum-space dispersion is the singularity in the density of states, leading to atomic-like discretization of the energy levels reminiscent of the Landau levels in the quantum Hall regime. If the Fermi level ($E_F$) of the material is tuned to lie at these singularities, the system can prefer to reduce the total electronic ground state energy *via* an energy gap opening that triggers the emergence of exotic quantum phase transitions. Thus, efforts to engineer flat bands around $E_F$ remains to be an active research focus where major advances are being made in various lattice systems, including in the moiré superlattices with gapless and gapped Dirac Hamiltonians.

Graphitic systems are among the family of materials that can host flat electronic bands,[11-17] with reports of high density of states near van Hove singularities at high binding energies and reduction of band dispersion in rhombohedral multilayers near charge neutrality. More recently, tBLG has emerged as a promising system due to its excellent tunability: in the degree of interlayer hybridization with twist angle[18-20] and in the possibility of using *in-situ* electrostatic gating for adjusting $E_F$ to occupy the flat moiré minibands or to achieve commensurate filling. The quenching of quasiparticle kinetic energy following the reduced miniband bandwidth at the magic angle is conducive to the emergence of interaction-driven phase transition. Partial fillings of the flat miniband in tBLG has resulted in the observation of correlated insulator[3], superconductivity[4,21,22], and orbital magnetism[22]. Richer physics may also arise by accounting for the sample environment, such as the ferromagnetism in magic angle tBLG with alignment to the hBN substrate[23,24].

Here we visualize the weak dispersion of the flat moiré miniband in tBLG near the magic angle twist with nanoARPES. ARPES provides a unique capability to resolve the $k$-space dispersion of the flatband. The high spatial resolution of nanoARPES enabled by the capillary focusing (~1 μm beam spot size, see method) is beneficial to counteract local structural inhomogeneity within the sample. The flat band is present even at room temperature near $E_F$ around the original $\bar{K}$ points of the constituent graphene monolayers.

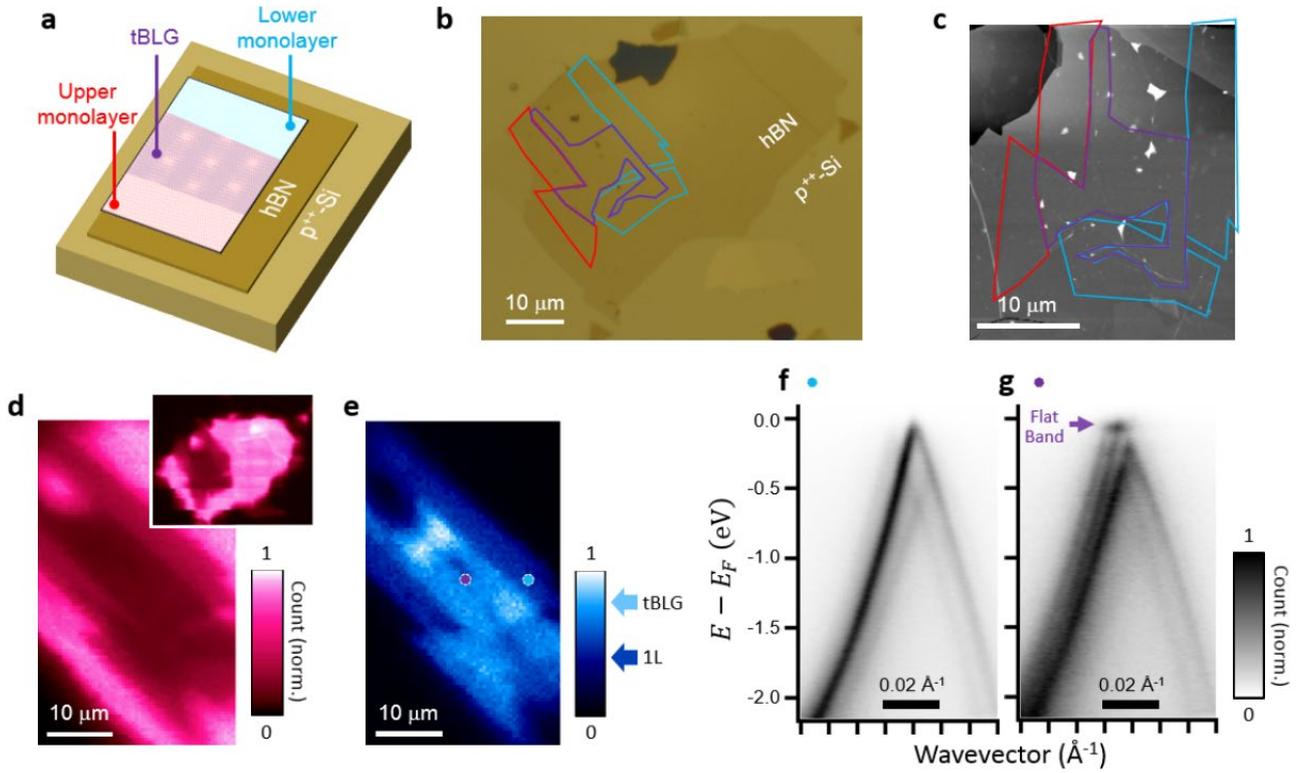

**Figure 1 | tBLG near magic angle twist on hBN substrate. a,** The schematic illustration of the tBLG/hBN/doped Si sample. **b,** Bright field optical micrograph and **c,** AFM image of the sample. The boundary of each segment is indicated as blue: lower graphene monolayer, red: upper graphene monolayer, purple: tBLG area. **d,e** SPEM image generated from real space mapping of the valence band spectra and integrating (d) the signal of hBN, and (e) the signal of graphene. A larger area scan covering the entire hBN flake is given as the inset in (d). The approximate intensity corresponding to monolayer and tBLG is marked at the color scale of (e). **f,g** Energy-momentum band dispersion around the $\overline{K}$ point of (f) lower graphene monolayer, and (g) tBLG with local twist near magic angle. These spectra were measured at the location marked with blue and purple circles in (e), respectively. The flat band in (g) is indicated with a purple arrow.

Fig. 1a shows the configuration of the sample. The sample was fabricated *via* a tear-and-stack method for controlled twist angle combined with stack inversion (see methods). This method allows production of uncapped tBLG on a flat hBN flake to minimize surface roughness and thus achieve improved momentum resolution during the photoemission spectroscopy. Fig. 1b is an optical image of such graphene sample on hBN, where the borders of the isolated monolayers and the tBLG region are indicated as determined from the atomic force microscopy (AFM) in Fig. 1c. Scanning photoemission microscopy (SPEM; Fig. 1d,e) that maps the real space distribution of generated photoelectrons from the valence band of graphene and hBN confirms the sample configuration. The intensity of the signal arising from graphene correlates with the layer number: the middle region has a stronger signal than the two sandwiching regions, matching the expected location of the monolayer and tBLG segments as demarcated in Fig. 1b. Similarly, the reduction of the hBN signal also correlates qualitatively with the attenuation from the different graphene thickness at each segment.

Fig. 1f,g displays the ARPES spectra collected from the lower graphene monolayer and the tBLG from locations as indicated in Fig. 1e. The monolayer spectrum exhibits the typical linear band dispersion characteristic for graphene and the sample is approximately charge neutral as the Dirac point is situated close to the Fermi level. Faint Dirac cone replicas are also seen surrounding the primary cone, which is a result of the superlattice periodicity between graphene and hBN[25] (further details are discussed in Supplementary Fig. 4). On the other hand, the spectrum from the tBLG area exhibits a sharp feature near the Fermi level corresponding to the flat electronic band. Subsequent discussions will be focused on the tBLG data (microscopy in Fig. 2 and ARPES in Fig. 3) that are collected from this location.

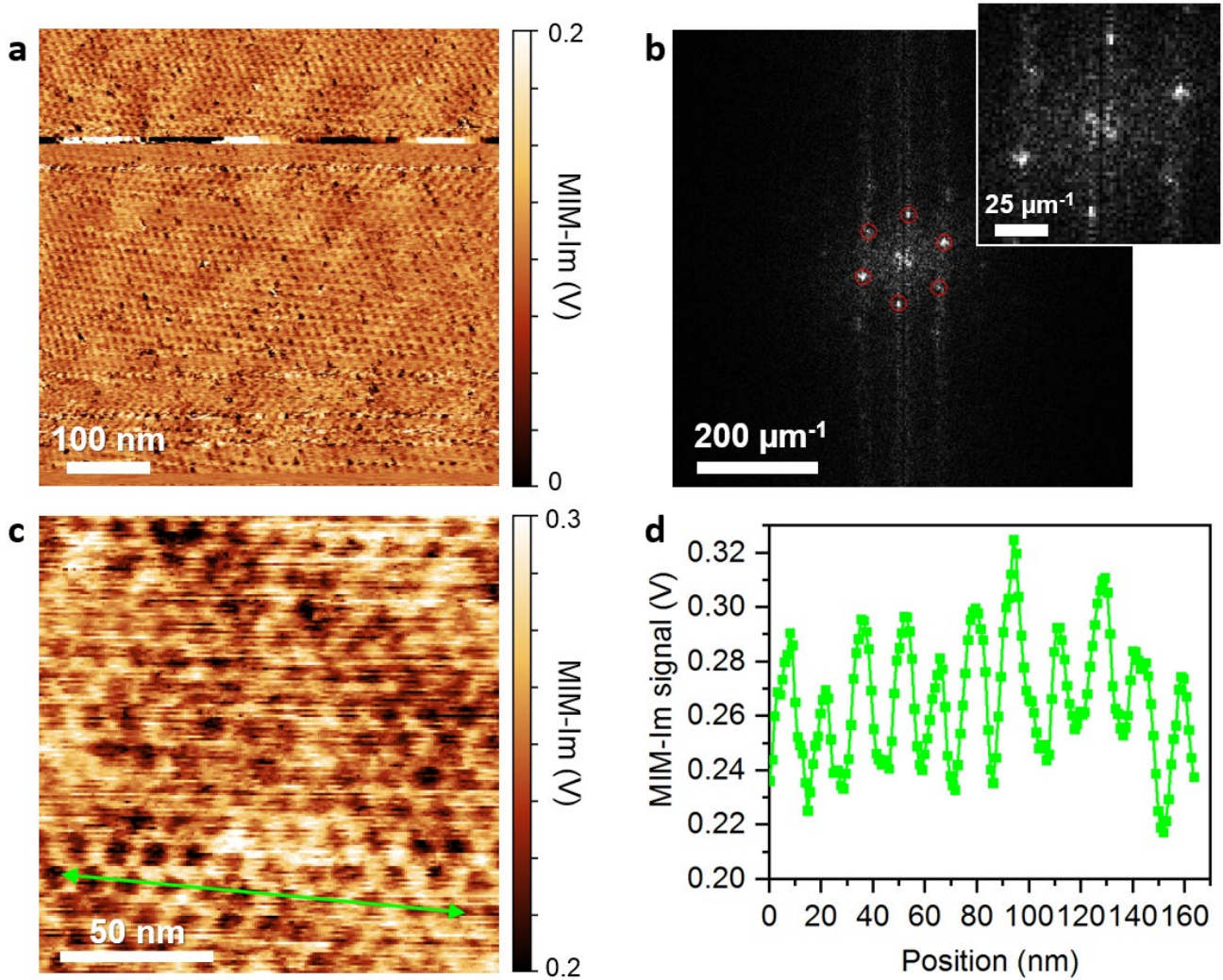

**Figure 2 | Moiré pattern visualized with MIM from tBLG in the location with flat electronic band feature. a,** Wide area map of MIM-Im signal. The periodic moiré pattern is clearly visible. **b,** The FFT of the image in (a) after excluding the horizontal scan artefact. The red circles marked the position of first-order spots from moiré period due to the graphene/graphene twist. *Inset*: Zoom-in of the area around the first-order spots. **c,** A magnified view of the moiré pattern. **d,** The MIM-Im signal profile taken along the green arrow in (c).

It is important to fabricate the twist angle of the tBLG close to the first magic angle (~1°) as previous ARPES measurements on tBLG with other much smaller[20] or larger[19,26,27] twist angles did not observe any flat band signatures. Also, the twist angle in tBLG is sensitive to disorder, strain, and temperature treatment that may alter the local twist angle from the intended design[5]. To confirm the local twist angle at tBLG sites where the flat band was observed in nanoARPES, we performed microwave impedance microscopy (MIM) imaging *after* completing the nanoARPES measurements (Fig. 2, also see Supplementary Fig. 3). The imaginary part of the complex microwave response (MIM-Im) in general increases monotonically as a function of the local sample conductivity[28]. Therefore, MIM-Im serves as a viable means to probe the moiré superlattice as an alternative to topographic imaging (Supplementary Fig. 2) and to conductive AFM mode while not requiring grounding electrodes.

Fig. 2a shows the real space MIM-Im map of the tBLG location with flat band feature. The MIM-Im signal presents a periodic modulation with six-fold rotational symmetry from the moiré pattern, a result of the misalignment of the constituent graphene monolayers. From the fast Fourier transform (FFT) of the image (Fig. 2b), we can deduce an averaged real-space periodicity of (14.7±0.4) nm, corresponding to a graphene/graphene twist angle of (0.96±0.03)°. Such periodicity can also be well-resolved directly in the line profile of the signal (Fig. 2c,d).

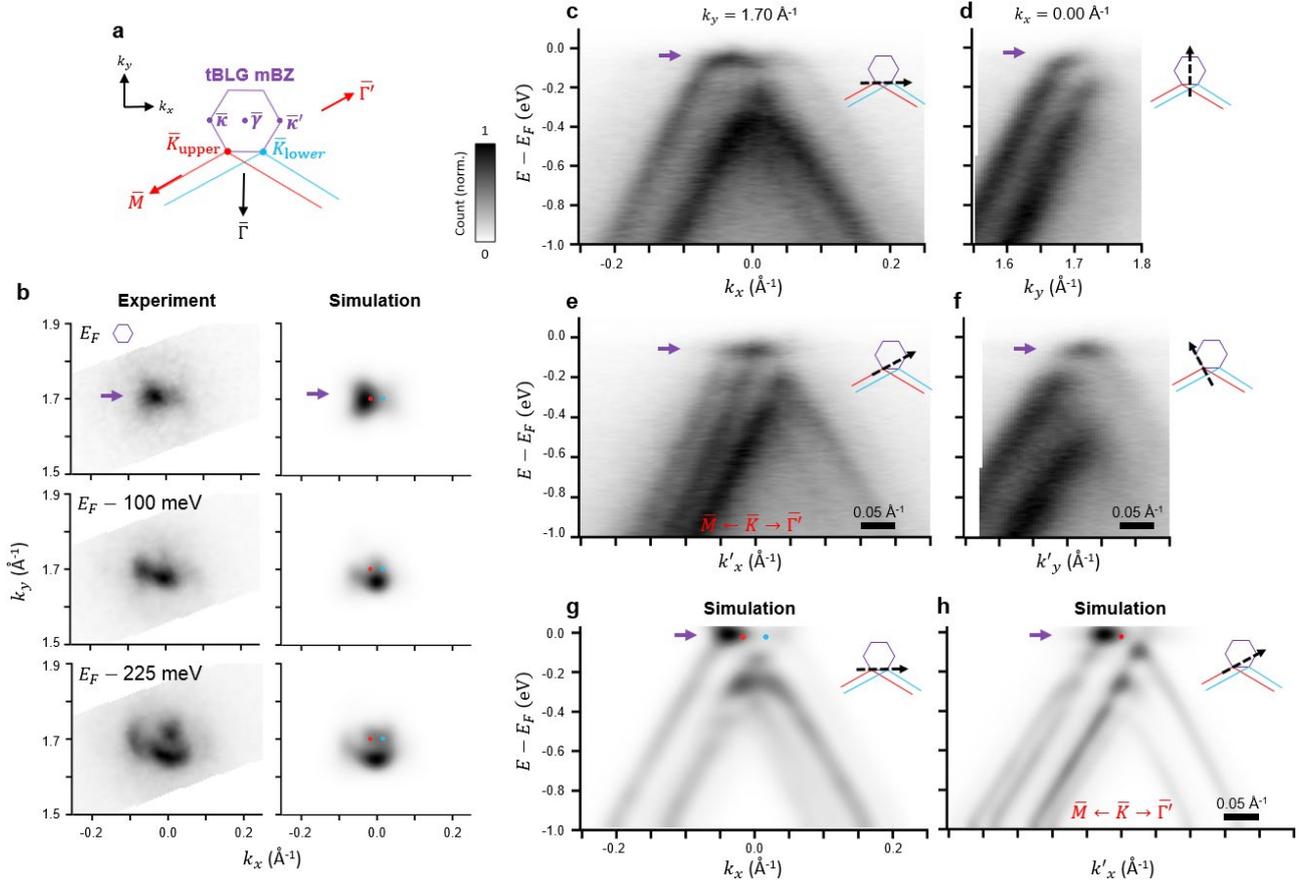

**Figure 3 | Electronic structure of tBLG and the visualization of the flat band dispersion.** The flat electronic band is marked with purple arrows. **a,** The geometry of the tBLG mBZ (purple hexagon) relative to the BZ of the original monolayers around the $\overline{K}$ point. The red and blue dots, as also superimposed in the simulation results in panels (b), (g), and (h), mark the location of the $\overline{K}$ points of the original monolayer constituents. The origin of axis $k_y$ is chosen to be along the $\overline{\Gamma} - \overline{\gamma}$ line. **b,** Equal energy cut for several binding energies. Images shown in the left and the right columns are the experimental and the simulation results, respectively. The size of the mBZ corresponding to the twist angle as determined in Fig. 2 is represented by the purple hexagon in the $E = E_F$ image. **c,** The dispersion of the tBLG band along $k_x$ at $k_y = 1.70$ Å$^{-1}$. **d,** The dispersion of the tBLG band along $k_y$ at $k_x = 0.00$ Å$^{-1}$. **e,** Alternative view of the tBLG band along an axis $k'_x$ that coincides with the path $\overline{M} - \overline{K} - \overline{\Gamma}'$ from the BZ of the upper monolayer. Here, $\overline{\Gamma}'$ is the center of the nearest-neighboring BZ. **f,** Band dispersion along an axis that is orthogonal to that in (e). **g,h,** Simulated spectral function of tBLG near the $\overline{K}$ point after accounting for lattice reconstruction. In (c)-(h), the direction of the momentum cut is illustrated in the inset of each panel. The experiment and simulated images follow the colorscale shown near (c).

Fig. 3 shows the photoemission spectra of the tBLG around the $\overline{K}$ points of the original Brillouin zone (BZ) from the constituent monolayers. The flat electronic bands are present near $E_F$, as also visualized experimentally along the various momentum cuts of the spectra in Fig. 3c-f (the cutting geometries are shown as the inset in each panel following the schematic in Fig. 3a). At the Fermi surface, the signal from the flat band is distributed around two intensity centers (Fig. 3b, top left). These distribution may originate from the states near the $\overline{K}$ point of the constituent upper and lower monolayer graphene, where the features from the lower monolayer appear to be dimmer due to photoelectron attenuation owing[19]. The flat band subtends over a width of ~0.1 Å$^{-1}$ along $k_x$, which is around twice larger than the $\overline{\kappa} - \overline{\gamma} - \overline{\kappa}'$ width of the tBLG mini BZ (mBZ, see Fig. 3a). The wide momentum extent of the flat band is consistent with the strong real space localization of carriers occupying these states. Such extendedness of the flat band states, along with the small outgoing branches away from the $\overline{K}$ points, can also be reproduced in the *ab initio*-informed tight-binding simulation of the spectral function (Fig. 3b, top right).

At higher binding energies, the equal energy cuts and the band dispersions also reveal the emergence of multiple Dirac cones instead of just two cones, which is what would have been observed from a Bernal-stacked bilayer graphene. This observation indicates the hybridization of the Dirac cones of the two monolayers due to a strong interlayer coupling of the constituent monolayers, as is expected for small twist angles and the periodic repetition of the moiré mBZ. Moreover, the periodic potential from the tBLG moiré superlattice is responsible for the avoided gaps that further split the bands[19]. These features are similar to those observed in other tBLG measurements[19,27] although in our sample the hybridized bands formed at lower binding energies due to the smaller twist angle.

Using the same choice of axes to Fig. 3c and 3e, we performed a simulation of the spectral function whose results are shown in Fig. 3g and 3h, respectively (see methods and the supplementary information for details). The simulated spectrum can reproduce qualitatively the salient features of the ARPES band dispersion, including the flat band at $E_F$ and the emergence of multiple Dirac cones.

In conclusion, we have visualized the presence of weakly dispersing flat electronic bands in tBLG near the magic twist angle. Our ARPES measurements provide a direct confirmation of the long-standing theoretical prediction for which only indirect signatures through transport or tunneling measurements were available. We have also demonstrated the capabilities of nanoARPES for accessing the electronic structure of moiré superlattice-induced flat bands in van der Waals heterostructures. It would be also of interest to perform detailed studies of nanoARPES in other graphitic moiré superlattice systems, including their behavior at different filling factors with in-situ electrostatic gating[29,30].

## Methods

**Fabrication and structural characterization.** We followed the tear-and-stack fabrication method to control the twist angle[31,32]. The step-by-step process is illustrated in Supplementary Fig. 1. We exfoliated a monolayer graphene and a 20 nm thick hBN flakes on a Si substrate with 285 nm thick $SiO_2$ film. We then used a stamp with polypropylene carbonate (PPC) coating. All pick-up processes are performed at 45°C. We first pick the hBN with the stamp and use the flake to pick part of the monolayer graphene. The substrate, mounted to a goniometer, is then rotated by an angle ~1.1°. The remaining monolayer segment was then picked-up to overlap with the monolayer already on the hBN, producing a tBLG. The PPC film is peeled from the stamp and transferred onto a clean highly doped Si substrate. Finally, the sample was annealed in vacuum at 250°C for 3 hours to remove the PPC film.

**Scanning probe microscopy.** The MIM characterization was performed with a modified Asylum MFP-3D AFM with commercial ScanWave electronics and coaxially shielded probes (PrimeNano Inc)[33]. Microwave frequency of 2.9 GHz was sent to the tip followed by the collection of the reflected signals. MIM signal is measured concurrently with the topographic imaging. The MIM measurement was conducted in contact mode with sub-10 nm lateral resolution under ambient condition. Additional AFM imaging to measure the topography was also performed with Asylum Cypher VRS and Park NX-10.

**NanoARPES measurement.** The sample was transported to the Microscopic and Electronic Structure Observatory (MAESTRO) at the Advanced Light Source where they were inserted in the nanoARPES UHV endstation with a base pressure better than $5\times10^{-11}$ mbar. The samples were annealed at 150°C for 24 hours prior to measurements in order to desorb adsorbates. The nanoARPES were measured with capillary focusing using the photon energy of 95 eV (except for Supplementary Fig. 4a-d which were measured at 147 eV) and a beam spot size of approximately 1 μm. The data were collected using a hemispherical Scienta R4000 electron analyzer. The net energy resolution of the in the nanoARPES data were around 34 meV. All measurements were carried out at room temperature.

**Spectral function simulation.** The unfolded band structure of tBLG was calculated using a tight-binding model based on a Wannier transformation procedure of DFT extracted electronic states of the lattice structure. A magic angle twist of 1.12° was used in the simulation. The lattice structure includes the effect of reconstruction as obtained from molecular dynamic simulations using a force-field parametrized to high-level exact exchange and random phase approximation (EXX+RPA) data[34,35]. We used an effective nearest neighbor hopping energy of $|t_0| \sim 3.1$ eV and Fermi velocity of $1.05\times10^6$ m/s within the F2G2 model[36] for the intralayer terms in the single layer graphene. The spectral function calculation follows the theory and band-unfolding scheme as described in Ref. 37, with the spectrum projected to the BZ of the lower graphene monolayer. To reproduce the asymmetry in the spectral function intensity owing to the upper and lower monolayer graphene, we assign an intensity weight ratio of 2:1 in the calculation. The equal energy cut is displayed by incorporating post-processing Gaussian broadening of the momentum space and cutoff energy of ±20 meV around the Fermi level in energy space.

## Data availability
The data that support the plots within this paper and other findings of this study are available from the corresponding author upon reasonable request. Simulation parameters are provided in the supplementary material and can be used as is in LAMMPS or with the KIM MD potential database.

## Acknowledgements
We thank S. Kahn for technical assistance in the sample fabrication setup. This work was supported as part of the Center for Novel Pathways to Quantum Coherence in Materials, an Energy Frontier Research Center funded by the U.S. Department of Energy, Office of Science, Basic Energy Sciences. The Advanced Light Source is supported by the Director, Office of Science, Office of Basic Energy Sciences, of the U.S. Department of Energy under Contract No. DE-AC02-05CH11231. The MIM measurements are supported by the Materials Sciences and Engineering Division of the U.S. Department of Energy under Contract No. DE-AC02-05-CH11231 (sp2-Bonded Materials Program KC2207). The MIM instrument at the Molecular Foundry were supported by the Office of Science, Office of Basic Energy Sciences, of the US Department of Energy under contract number DE-AC02-05CH11231. K.W. and T.T. acknowledge support from the Elemental Strategy Initiative, conducted by the MEXT, Japan and CREST (JPMJCR15F3), JST. N.L. acknowledges the Korean national Research Foundation grant NRF-2018R1C1B6004437, the Korea Research Fellowship Program funded by the Ministry of Science and ICT (KRF-2016H1D3A1023826), as well as the computational resources by KISTI through grant KSC-2018-CHA-0077. J.J. was supported by the Samsung Science and Technology Foundation under project SSTF-BAA1802-06.

## Author contribution
F.W., E.R., and M.I.B.U. conceived the project. M.I.B.U. developed the sample preparation method and fabricated all samples with J.Z. assistance. R.J.K., A.B., and E.R performed the nanoARPES experiments. R.J.K., C.J., A.B., and E.R. developed and maintained the nanoARPES setup. M.I.B.U. and F.W. analyzed the nanoARPES experimental data with A.B. and E.R. providing guidance. K.L. contributed in the MIM instrumentation setup. K.L. and M.I.B.U. performed AFM and MIM, and analyzed the data with P.D.A., A.W.B., and A.Z. providing guidance. J.J. and N.L. calculated the spectral functions. H.L. and S.Z. contributed in surface cleaning process. L.J. performed SNOM. K.W. and T.T. grew the hBN single crystal. F.W. and E.R. supervised the project. M.I.B.U. and F.W. wrote the manuscript with inputs from all authors.

## Financial and non-financial competing interests
The authors declare no competing interests.

## Additional information
**Correspondence and requests for materials** should be addressed to F.W., A.B., and J.J.